\documentclass[twocolumn,amsmath,amssymb]{revtex4}

\usepackage{graphicx}
\usepackage{amsmath}
\usepackage{amssymb}

\begin{document}

\title{Doping and temperature dependence of electronic Raman
response in cuprate superconductors}

\author{Zhihao Geng and Shiping Feng}

\affiliation{Department of Physics, Beijing Normal University,
Beijing 100875, China}


\begin{abstract}
The doping and temperature dependence of the electronic Raman
response in cuprate superconductors is studied within the kinetic
energy driven superconducting mechanism. It is shown that the
temperature dependent depletion at low-energy shifts is faster in
the $B_{1g}$ symmetry than in the $B_{2g}$ symmetry. In analogy to
the domelike shape of the doping dependent superconducting
transition temperature, the maximal peak energy in the $B_{2g}$
channel occurs around the optimal doping, and then decreases in both
underdoped and overdoped regimes. Moreover, the overall density of
Cooper pairs increases with increasing doping in the underdoped
regime.
\end{abstract}

\maketitle

Cuprate superconductors are doped Mott insulators with the strong
electron correlation dominating the entire phase diagram
\cite{shen}. After over 20 years extensive studies, it has become
clear that superconductivity in doped cuprates results when
electrons pair up into Cooper pairs \cite{tsuei} as in the
conventional superconductors \cite{bcs}. However, the
superconducting (SC) transition temperature $T_{c}$ is strongly
dependent on doping, and takes a domelike shape with the underdoped
and overdoped regimes on each side of the optimal doping where
$T_{c}$ researches its maximum. Moreover, as a natural consequence
of the unconventional SC mechanism \cite{anderson}, the Cooper pairs
have a dominant d-wave symmetry \cite{tsuei,shen}. Since many
physical properties of the two-particle electron dynamics have been
attributed to particular characteristics of low energy excitations
determined by the electronic Raman response (ERR) \cite{t1,d1}, in
this case, a central issue to clarify the nature of the electron
dynamics is how ERR evolves with doping and temperature.

Experimentally, by virtue of systematic studies using the ERR
measurement technique, some essential features of the evolution of
the two-particle excitations in cuprate superconductors with doping
and temperature in the SC state have been established
\cite{t1,d1,e1,e2,e3,e4,e5,e6,e7,e8,e9,e10}: (a) the $B_{1g}$
orientation projects out excitations around the antinodal region in
the Brillouin zone where the energy gap is maximal; (b) the $B_{2g}$
orientation projects out excitations around the nodal region where
the energy gap vanishes; (c) as a consequence of both cases (a) and
(b), ERR has a peak at the antinodes in the $B_{1g}$ symmetry and at
slightly lower energy at the nodes in the $B_{2g}$ symmetry;
however, (d) in the underdoped regime, the peak in the $B_{1g}$
symmetry shifts to a higher energy with decreasing doping and tracks
of the antinodal gap, while the peak in the $B_{2g}$ symmetry shows
the same doping dependence as $T_{c}$ and shifts to a lower energy
with decreasing doping; and (e) the $B_{2g}$ response depends
linearly on energy $\omega$ in the limit $\omega\sim 0$ for the
energy gap vanishing along with the diagonal directions of the
Brillouin zone, while for the $B_{1g}$ orientation, the Raman vertex
vanishes along with the energy gap at the same directions, which
yields an additional $\omega^{2}$ contribution from the line nodes
of the vertex, and the resulting ERR varies as $\omega^{3}$.
Theoretically, an agreement has emerged that a simple d-wave
Bardeen-Cooper-Schrieffer (BCS) formalism is useful in the
phenomenological description of ERR in cuprate superconductors
\cite{d1,devereaux94,manske97,branch95,dahm98,chubukov06}.  In
particular, incorporating the d-wave BCS formalism with the
spin-fluctuation model \cite{branch95,dahm98,chubukov06}, the role
of the collective modes in the ERR experiment on cuprate
superconductors is studied \cite{dahm98}, and the result shows that
the collective modes of the amplitude fluctuation of the d-wave gap
yield a broad peak above the threshold in the $B_{1g}$ spectrum.
Moreover, it has been shown that ERR in the $B_{1g}$ symmetry allows
one to distinguish between phonon-mediated and magnetically mediated
d-wave superconductivity \cite{chubukov06}. However, to the best of
our knowledge, ERR in cuprate superconductors has not been treated
starting from a microscopic SC theory, and no explicit calculation
of the doping dependence of ERR has been made so far. Very recently,
using a simple relationship between the integrated ERR and density
of Cooper pairs (DOCP), it is shown experimentally \cite{blanc} that
in a similar way to the superfluid density \cite{uemura89}, the
overall DOCP strongly increases with increasing doping in the
underdoped regime \cite{blanc}. Thus this recently discovered doping
dependence of the overall DOCP is also calling for an explanation.

In this Letter, we study ERR in cuprate superconductors based on the
kinetic energy (KE) driven SC mechanism \cite{feng03}. We evaluate
explicitly the ERR function in terms of the Raman density-density
correlation function (RDDCF), and qualitatively reproduce some main
features of ERR in cuprate superconductors
\cite{e1,e2,e3,e4,e5,e6,e7,e8,e9,e10,blanc}. In particular, we show
that the temperature dependent depletion at low-energy shifts is
faster in the $B_{1g}$ symmetry than in the $B_{2g}$ symmetry.
Moreover, in analogy to the domelike shape of the doping dependence
of $T_{c}$, the maximal peak energy in the $B_{2g}$ channel occurs
around the optimal doping, and then decreases in both underdoped and
overdoped regimes. Furthermore, the overall DOCP increases with
increasing doping in the underdoped regime.

We start from the $t$-$J$ model \cite{anderson,shen},
\begin{eqnarray}\label{tjham}
H&=&-t\sum_{i\hat{\eta}\sigma}C_{i\sigma}^{\dagger}
C_{i+\hat{\eta}\sigma}+t'\sum_{i\hat{\eta}'\sigma}C_{i\sigma}^{\dagger}
C_{i+\hat{\eta}'\sigma}\nonumber\\
&+&\mu\sum_{i\sigma}C_{i\sigma}^{\dagger}
C_{i\sigma}+J\sum_{i\hat{\eta}}S_{i}\cdot S_{i+\hat{\eta}},
\end{eqnarray}
acting on the Hilbert subspace with no doubly occupied site, i.e.,
$\sum_{\sigma}C_{i\sigma}^{\dagger}C_{i\sigma}\le 1 $, where
$\hat{\eta}=\pm\hat{x},\pm\hat{y}$, $\hat{\eta}'=\pm\hat{x}\pm
\hat{y}$, $C_{i\sigma}^{\dagger}(C_{i\sigma})$ is the creation
(annihilation) operator of an electron with spin $\sigma$, ${\bf
S}_{i}=(S_{i}^{x},S_{i}^{y},S_{i}^{z})$ are spin operators, and
$\mu$ is the chemical potential. To deal with the constraint of no
double occupancy, the charge-spin separation (CSS) fermion-spin
theory \cite{feng04,feng08} has been developed, where the physics of
no double occupancy is taken into account by representing the
constrained electron operators as $C_{i\uparrow}=
h_{i\uparrow}^{\dagger}S_{i}^{-}$ and $C_{i\downarrow}=
h_{i\downarrow}^{\dagger}S_{i}^{+}$, with the spinful fermion
operator $h_{i\sigma}=e^{-i\Phi_{i\sigma}}h_{i}$ that describes the
charge degree of freedom together with some effects of spin
configuration rearrangements due to the presence of the doped hole
itself, while the spin operator $S_{i}$ represents the spin degree
of freedom, then the electron local constraint for the single
occupancy is satisfied in analytical calculations. In this CSS
fermion-spin representation, the $t$-$J$ model (\ref{tjham}) is
expressed as,
\begin{eqnarray}\label{cssham}
H&=&t\sum_{i\hat{\eta}}(h^{\dagger}_{i+\hat{\eta}\uparrow}h_{i\uparrow}
S_{i}^{+}S_{i+\hat{\eta}}^{-}+h_{i+\hat{\eta}\downarrow}^{\dagger}
h_{i\downarrow}S_{i}^{-}S_{i+\hat{\eta}}^{+})\nonumber\\
&-&t'\sum_{i\hat{\eta}'}
(h_{i+\hat{\eta}'\uparrow}^{\dagger}h_{i\uparrow}S_{i}^{+}
S_{i+\hat{\eta}'}^{-}+h_{i+\hat{\eta}'\downarrow}^{\dagger}
h_{i\downarrow}S_{i}^{-}S_{i+\hat{\eta}'}^{+})\nonumber\\
&-&\mu\sum_{i\sigma}h_{i\sigma}^{\dagger}h_{i\sigma}+J_{{\rm eff}}
\sum_{i\hat{\eta}}S_{i}\cdot S_{i+\hat{\eta}},
\end{eqnarray}
with $J_{{\rm eff}}=(1-\delta)^{2}J$, and $\delta=\langle
h_{i\sigma}^{\dagger}h_{i\sigma}\rangle=\langle h_{i}^{\dagger}
h_{i}\rangle$ is the hole doping concentration.

For an understanding of the physical properties of cuprate
superconductors in the SC state, the KE driven SC mechanism has been
developed \cite{feng03} based on the CSS fermion-spin theory, where
the interaction between charge carriers and spins from the KE term
in the $t$-$J$ model (\ref{cssham}) induces the d-wave charge
carrier pairing state by exchanging spin excitations, then the
electron Cooper pairs originating from the charge carrier pairing
state are due to the charge-spin recombination, and their
condensation reveals the SC ground-state. In particular, this SC
state is a conventional BCS-like with the d-wave symmetry
\cite{guo07,feng08}, so that the basic d-wave BCS formalism is still
valid in discussions of the low energy electron quasiparticle
excitations of cuprate superconductors, although the pairing
mechanism is driven by KE by exchanging spin excitations. Following
these previous discussions \cite{guo07,feng03}, the full charge
carrier Green function can be obtained in the Nambu representation
as,
\begin{eqnarray}
g({\bf{k}},i\omega_n)=Z_{\rm{hF}}\,\frac{i\omega_n\tau_0 +
\bar{\xi}_{\bf{k}}\tau_3 - \bar{\Delta}_{\rm{hZ}}({\bf{k}})
\tau_1}{(i\omega_n)^2 - E_{{\rm{h}}{\bf{k}}}^2},
\label{holegreenfunction}
\end{eqnarray}
where $\tau_{0}$ is the unit matrix, $\tau_{1}$ and $\tau_{3}$ are
Pauli matrices, the charge carrier quasiparticle spectrum $E_{h{\bf
k}}=\sqrt {\bar{\xi^{2}_{{\bf k}}}+\mid \bar{\Delta}_{hZ}({\bf k})
\mid^{2}}$ with the renormalized d-wave charge carrier pair gap
function $\bar{\Delta}_{hZ}({\bf k})=\bar{\Delta}_{hZ}[{\rm cos}
k_{x}-{\rm cos}k_{y}]/2$, while the charge carrier quasiparticle
coherent weight $Z_{hF}$ and other notations are defined as same as
in Ref. \cite{guo07}, and have been determined by the
self-consistent calculation \cite{guo07,feng03}.

In the CSS fermion-spin representation \cite{feng04}, the electron
Green's function,
\begin{eqnarray}
G({\bf k}, i\omega_n)=\left(
\begin{array}{cccc}
G_{11}({\bf k},i\omega_n), & G_{12}({\bf k},i\omega_n) \\
G_{21}({\bf k},i\omega_n), & G_{22}({\bf k},i\omega_n)
\end{array} \right) \,, \label{electrongreenfunction}
\end{eqnarray}
is a convolution of the spin Green's function and charge carrier
Green's function (\ref{holegreenfunction}), and its diagonal and
off-diagonal components $G_{11} (i-j,t-t') =\langle\langle
C_{i\sigma}(t); C^{\dagger}_{j\sigma} (t')\rangle \rangle$ and
$G_{21}(i-j,t-t')=\langle \langle C^{\dagger}_{i\uparrow}(t);
C^{\dagger}_{j\downarrow}(t')\rangle \rangle$ have been given in
Ref. \cite{guo07}.

The ERR function $\tilde{S}({\bf q},\omega)$ is obtained from the
imaginary part of RDDCF $\tilde{\chi}({\bf q},\omega)$ as \cite{d1},
\begin{eqnarray}
\tilde{S}({\bf q},\omega)=-{1\over\pi}[1+n_{B}(\omega)]{\rm Im}
\tilde{\chi}({\bf q},\omega), \label{ramanscattering}
\end{eqnarray}
where $n_{B}(\omega)$ is the boson distribution function, while
RDDCF $\tilde{\chi}({\bf q},\omega)$ is defined as,
\begin{eqnarray}
\tilde{\chi}({\bf q},\tau-\tau')=-\langle T\rho_{\gamma}({\bf q},
\tau) \rho_{\gamma}(-{\bf q},\tau')]\rangle,
\end{eqnarray}
where $\rho_{\gamma}({\bf q})=\sum_{{\bf k}}\gamma_{\bf k}
C^{\dagger}_{{\bf k}+{{\bf q}\over 2}}\tau_{3} C_{{\bf k}-{{\bf q}
\over 2}}$ is the Raman density operator in the Nambu
representation, with the bare Raman vertex $\gamma_{\bf k}$ has been
classified by the representations $B_{1g}$, $B_{2g}$, and $A_{1g}$
of the point group $D_{4h}$ as \cite{d1,manske97},
\begin{eqnarray}
\gamma_{\bf k}=\left\{
\begin{array}{ll}
b_{\omega_{i},\omega_{s}}\left[\cos(k_{x}a)-\cos(k_{y}a)
\right]/4,& B_{1g},\\
b'_{\omega_{i},\omega_{s}}\sin(k_{x}a)\sin(k_{y}a),& B_{2g},\\
a_{\omega_{i},\omega_{s}}\left[\cos(k_{x}a)+\cos(k_{y}a)\right]/4,&
A_{1g},
\end{array}
\right.
\label{vertex}
\end{eqnarray}
respectively, where as a qualitative discussion, the magnitude of
the energy dependence of the prefactors $b$, $b'$ and $a$ can be
rescaled to units. In this case, $\tilde{\chi}({\bf q},\omega)$ can
be obtained in terms of the electron Green's function
(\ref{electrongreenfunction}) as,
\begin{widetext}
\begin{eqnarray}
&\tilde{\chi}_{\gamma_{1}\gamma_{2}}&({\bf q},\omega)={1\over 2}
Z^{2}_{F}{1\over N^{3}}\sum_{{\bf k},{\bf p},{\bf p'}}\gamma_{1{\bf
k}}\gamma_{2{\bf k}}{B_{{\bf p}}B_{{\bf p'}}\over \omega_{{\bf p}}
\omega_{{\bf p'}}}\sum_{\mu\nu\mu'\nu'=1,2} L_{\mu}({\bf p},{\bf p}
+{\bf k}+{\bf q})L_{\mu'}({\bf p'},{\bf p'}+{\bf k})\left
[U^{2}_{\nu h{\bf p}+{\bf k}+{\bf q}}U^{2}_{\nu' h{\bf p'}+{\bf k}}
\right .\nonumber\\
&-&\left . (-1)^{\nu+\nu'}{\bar{\Delta}_{hZ}({\bf p}+{\bf k}+{\bf
q})\bar{\Delta}_{hZ}({\bf p'}+{\bf k})\over 4E_{h{\bf p}+ {\bf k}
+{\bf q}}E_{h{\bf p'}+{\bf k}}}\right ]{n_{F}[(-1)^{\nu'}E_{h{\bf
p'}+{\bf k}}-(-1)^{\mu'+\nu'} \omega_{\bf p'}]-
n_{F}[(-1)^{\nu}E_{h{\bf p}+{\bf k}+{\bf q}}-
(-1)^{\mu+\nu}\omega_{\bf p}]\over \omega+(-1)^{\nu'}E_{h{\bf p'}+
{\bf k}}-(-1)^{\mu'+\nu'}\omega_{\bf p'}-(-1)^{\nu}E_{h{\bf p}+{\bf
k}+{\bf q}}-(-1)^{\mu+\nu}\omega_{\bf p}},~~~~
\label{ramancorrelationfunction}
\end{eqnarray}
\end{widetext}
where the electron quasiparticle coherent weight $Z_{F}=Z_{hF}/2$,
$n_{F}(\omega)$ is the fermion distribution function, $U^{2}_{1h{\bf
k}}=(1+\bar{\xi_{{\bf k}}}/E_{h{\bf k}})/2$ and $U^{2}_{2h{\bf k}}
=(1-\bar{\xi_{{\bf k}}}/E_{h{\bf k}})/2$, while the spin excitation
spectrum $\omega_{{\bf p}}$ and $B_{\bf p}$ have been given in Ref.
\cite{guo07}.

\begin{figure}[h!]
\includegraphics[scale=0.3]{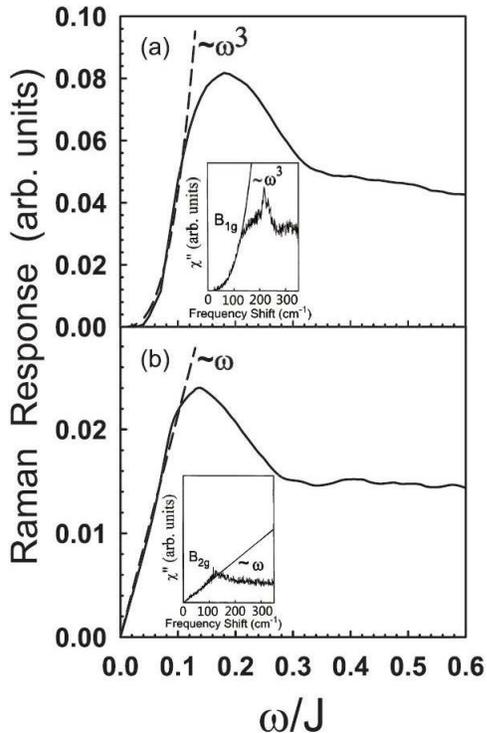}
\caption{(a) $B_{1g}$ and (b) $B_{2g}$ spectra as a function of
energy at $p=0.15$ for $T=0.002J$. The dashed lines are a cubic and
a linear fit for the low-energy $B_{1g}$ and $B_{2g}$ spectra,
respectively. Inset: the corresponding experimental results taken
from Ref. \onlinecite{e3}.} \label{fig1}
\end{figure}

In cuprate superconductors, although the values of $J$ and $t$ are
believed to vary somewhat from compound to compound \cite{shen},
however, as a qualitative discussion, the commonly used parameters
in this paper are chosen as $t/J=2.5$ and $t'/t=0.3$. In this case,
we have performed a calculation for the ERR function
(\ref{ramanscattering}) in both $B_{1g}$ and $B_{2g}$ orientations,
and the results of (a) the $B_{1g}$ and (b) $B_{2g}$ spectra with
doping $p=0.15$ at temperature $T=0.002J$ are plotted in Fig.
\ref{fig1}. For comparison, the corresponding experimental results
\cite{e3} are also presented in Fig. \ref{fig1} (inset). Obviously,
both $B_{1g}$ and $B_{2g}$ spectra are characterized by the presence
of the pair-breaking peaks, however, the peak in the $B_{1g}$
geometry develops at roughly $30\%$ higher energy than that in the
$B_{2g}$ geometry. Moreover, we have also fitted our present
results, and found that the low-energy spectra almost rise as
$\omega^{3}$ in the $B_{1g}$ channel and linearly with $\omega$ in
the $B_{2g}$ channel. However, in contrast to the conventional
superconductors, the intensities of ERR in both $B_{1g}$ and
$B_{2g}$ channels are equal to zero when energy $\omega=0$, in other
words, there are no sharp onset of the intensities at a threshold.
This can be understood from the physical property of RDDCF
(\ref{ramancorrelationfunction}). In low temperatures, the spins
center around the $[\pm\pi,\pm\pi]$ points, then the main
contribution from the spins comes from the $[\pm\pi,\pm\pi]$ points.
In this case, RDDCF (\ref{ramancorrelationfunction}) in the momentum
transfers ${\bf q} \sim 0$ limit can be approximately reduced in
terms of $\omega_{{\bf p}=[\pm\pi,\pm\pi]}\sim 0$ and one of the
self-consistent equations $1/2=\langle S_{i}^{+} S_{i}^{-}\rangle
=(1/N)\sum_{{\bf p}}B_{{\bf p}}{\rm coth}(\beta \omega_{{\bf p}}/2)
/(2\omega_{{\bf p}})$ as \cite{guo07},
\begin{eqnarray}
\tilde{\chi}_{\gamma_{1}\gamma_{2}}({\bf q}\sim 0,\omega)&\approx&
Z^{2}_{F}{1\over N}\sum_{\bf k}\gamma_{1{\bf k}}\gamma_{2{\bf k}}
{\bar{\Delta}^{2}_{Z}({\bf k})\over E^{2}_{\bf k}}{\rm tanh}
[{1\over 2}\beta E_{\bf k}]\nonumber\\
&\times& \left ({1\over 2E_{\bf k}+\omega} +{1\over 2E_{\bf
k}-\omega}\right ), \label{ramancorrelationfunction1}
\end{eqnarray}
where the renormalized d-wave electron pair gap function
$\bar{\Delta}_{Z}({\bf k })\approx\bar{\Delta}_{hZ}({\bf k+k_{A}})$
with ${\bf k_{A}}= [\pi,\pi]$, and the electron quasiparticle
spectrum $E_{{\bf k}}\approx E_{h{\bf k+k_{A}}}$. Since the d-wave
gap function vanishes along the diagonal directions of the Brillouin
zone, then as seen from Eq. (\ref{ramancorrelationfunction1}), this
leads to an absence of the threshold in cuprate superconductors.
These results are qualitatively consistent with the experimental
results \cite{e1,e2,e3}.

\begin{figure}[h!]
\includegraphics[scale=0.3]{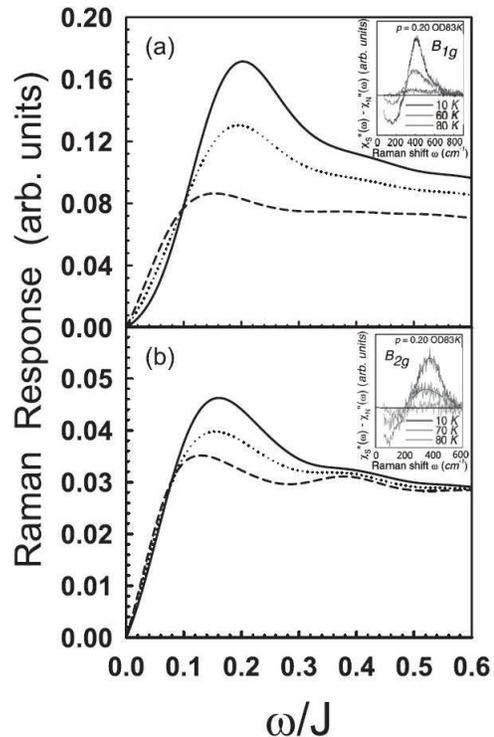}
\caption{(a) $B_{1g}$ and (b) $B_{2g}$ spectra as a function of
energy at $p=0.20$ for $T=0.03J$ (solid line), $T=0.05J$ (dotted
line), and $T=0.07J$ (dashed line). Inset: the corresponding
experimental results taken from Ref. \onlinecite{e10}.} \label{fig2}
\end{figure}

\begin{figure}[h!]
\includegraphics[scale=0.3]{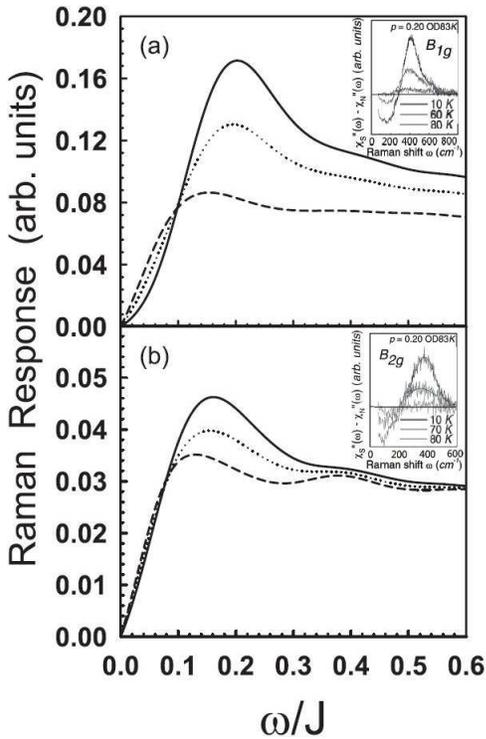}
\caption{(a) $B_{1g}$ and (b) $B_{2g}$ spectra as a function of
energy at $p=0.20$ (solid line) and $p=0.22$ (dotted line) for
$T=0.002J$. Inset: the corresponding experimental results taken from
Ref. \onlinecite{e6}.} \label{fig3}
\end{figure}

For a better understanding of the evolution of ERR with temperature,
we have further performed a calculation for the ERR function
(\ref{ramanscattering}) with different temperatures, and the results
of $\tilde{S}({\bf q}, \omega)$ in (a) the $B_{1g}$ and (b) $B_{2g}$
channels as a function of energy with $T=0.03J$ (solid line),
$T=0.05J$ (dotted line), and $T=0.07J$ (dashed line) for $p=0.20$
are plotted in Fig. \ref{fig2} in comparison with the corresponding
experimental results \cite{e10} (inset). Our results show that there
is a strong temperature dependence of the peaks in both $B_{1g}$ and
$B_{2g}$ orientations in the overdoped regime, where both $B_{1g}$
and $B_{2g}$ peaks at low temperatures soften in energy and
decreases in intensity as the temperature is raised. Moreover, as
seen from Eq. (\ref{ramancorrelationfunction1}), the intensity of
the peak follows a pair gap type temperature dependence, and
disappears at $T_{c}$. Simultaneously, the low energy continuum
grows in agreement with the transfer of spectral weight.
Furthermore, the temperature dependent depletion at low-energy
shifts is faster in the $B_{1g}$ symmetry than in the $B_{2g}$
symmetry. Although the strong temperature dependence of the $B_{2g}$
spectrum throughout the SC dome and $B_{1g}$ spectrum in the
overdoped regime are qualitatively consistent with the corresponding
experimental results \cite{e9,e10}, the temperature dependence of
the $B_{1g}$ spectrum in the underdoped regime is in disagreement
with the corresponding experimental result \cite{e9,e10}, where the
nontemperature dependence (or weakly increases as the temperature is
raised) is observed. This reflects that the excitations observed in
the $B_{1g}$ channel is only related to superconductivity in the
overdoped regime, and then ERR in the $B_{1g}$ and $B_{2g}$ channels
become drastically distinct in the underdoped regime, which can be
seen more clearly in the following discussions of the doping
dependence of ERR in the $B_{1g}$ channel.

\begin{figure}[h!]
\includegraphics[scale=0.25]{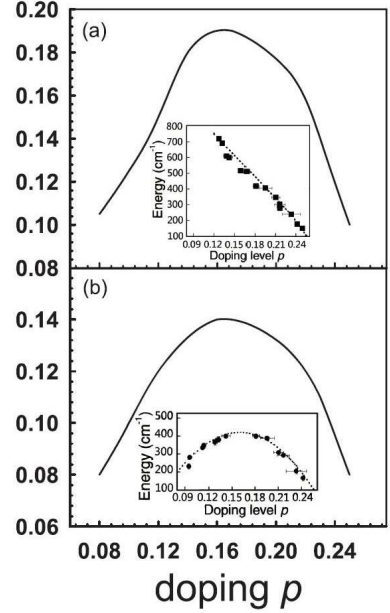}
\caption{(a) $B_{1g}$ and (b) $B_{2g}$ peaks as a function of doping
for $T=0.002J$. Inset: the corresponding experimental results taken
from Ref. \onlinecite{e9}.} \label{fig4}
\end{figure}

\begin{figure}[h!]
\includegraphics[scale=0.3]{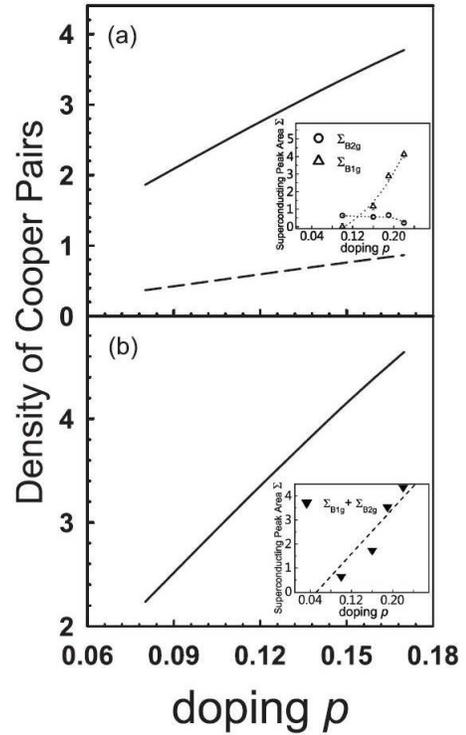}
\caption{Doping dependence of (a) the density of Cooper pairs in the
$B_{1g}$ (solid line) and $B_{2g}$ (dashed line) channels, and (b)
the sum of the densities of Cooper pairs in the $B_{1g}$ and
$B_{2g}$ channels for $T=0.002J$. Inset: the corresponding
experimental results taken from Ref. \onlinecite{blanc}.}
\label{fig5}
\end{figure}

To analyze the evolution of ERR in Fig. \ref{fig1} with doping, we
have made a series of calculations for the ERR function
(\ref{ramanscattering}) with different dopings, and the results of
(a) the $B_{1g}$ and (b) $B_{2g}$ spectra as a function of energy
with $p=0.20$ (solid line) and $p=0.22$ (dotted line) at $T=0.002J$
are plotted in Fig. \ref{fig3} in comparison with the corresponding
experimental results \cite{e6} (inset). It is shown clearly that
although the intensities of the peaks in both $B_{1g}$ and $B_{2g}$
symmetries are strongly suppressed in the overdoped regime when
reducing doping, the peak energies change drastically, i.e., the
peak energies in both $B_{1g}$ and $B_{2g}$ channels decrease with
increasing doping. Furthermore, we have found that the peak energies
in both $B_{1g}$ and $B_{2g}$ channels increase with increasing
doping in the underdoped regime. To show this point clearly, we have
calculated the ERR function (\ref{ramanscattering}) throughout the
SC dome, and then employed the shift of the leading-edge mid-point
as a measurement of the magnitude of the pair gap at each doping
just as it has been done in the ERR experiments
\cite{e4,e5,e6,e7,e8,e9}. The results for the extracted (a) $B_{1g}$
and (b) $B_{2g}$ peaks as a function of doping with $T=0.002J$ are
plotted in Fig. \ref{fig4}. For comparison, the corresponding
experimental results \cite{e9} are also presented in Fig. \ref{fig4}
(inset). The result of the peak energy in the $B_{2g}$ symmetry
continuously follows $T_{c}$ throughout the SC dome as
$\omega^{B_{2g}}_{{\rm peak}}\propto T_{c}$, i.e., the maximal peak
energy $\omega^{B_{2g}}_{{\rm peak}}$ in the $B_{2g}$ symmetry
occurs around the optimal doping, and then decreases in both
underdoped and overdoped regimes. In this sense, it can be
identified with the SC gap in the whole doping range, in qualitative
agreement with the experimental results \cite{e4,e5,e6,e7,e8,e9}.
However, in the $B_{1g}$ orientation, although the doping dependence
of the peak energy is qualitatively consistent with the experimental
data in the overdoped regime, it is in disagreement with the
experimental data in the underdoped regime \cite{e4,e5,e6,e7,e8,e9},
where the peak energy increases essentially linearly as doping is
reduced. This reflects that the peak energy in the $B_{1g}$ symmetry
is progressively disconnected from superconductivity as one goes
from the overdoped regime to the underdoped regime.

Now we turn to discuss the evolution of DOCP with doping. In the
case of the low temperatures and momentum transfers ${\bf q} \sim 0$
limit, the integral of the imaginary part of RDDCF over energy
$\omega$ can be obtained from Eq. (\ref{ramancorrelationfunction})
as,
\begin{eqnarray}
\rho_{{\rm pairs}}&=&\int^{\infty}_{0}d\omega
\tilde{\chi}_{\gamma_{1} \gamma_{2}}({\bf q} \sim 0,\omega)
\nonumber\\
&\approx& 2\pi Z^{2}_{F}{1\over N}\sum_{\bf k} \gamma_{1{\bf
k}}\gamma_{2{\bf k}} {\bar{\Delta}^{2}_{Z}({\bf k}) \over E^{2}_{\bf
k}}{\rm tanh} [{1\over 2}\beta E_{\bf k}]. ~~~~~
\label{cooperpair}
\end{eqnarray}
However, the sum $\sum_{\bf k}(\bar{\Delta}^{2}_{Z}({\bf k})/
E^{2}_{\bf k}){\rm tanh}(\beta E_{\bf k}/2)$ is equal to $4\sum_{\bf
k}(U_{\bf k}V_{\bf k})^{2}{\rm tanh}(\beta E_{\bf k}/2)$, where the
electron quasiparticle coherence factors $U^{2}_{{\bf k} }\approx
U^{2}_{2h{\bf k+k_{A}}}$ and $V^{2}_{{\bf k}}\approx U^{2}_{1h{\bf
k+k_{A}}}$ describe the probabilities of the Cooper pair being
occupied and unoccupied, respectively. Furthermore, this sum is not
vanishing only around the Fermi energy in the range of
$2\bar{\Delta}_{Z}({\bf k})$. This quantity corresponds to DOCP
\cite{blanc}, formed around the Fermi surface as the energy gap is
opening. In this sense, $\rho_{{\rm pairs}}$ is proportional to
DOCP, weighted by the Raman vertex \cite{blanc}. In this case, we
have performed a calculation for the doping dependence of
$\rho_{{\rm pairs}}$, and the results of (a) $\rho^{{\rm B_{1g}}}
_{{\rm pairs}}$ (solid line) and $\rho^{{\rm B_{2g}}}_{{\rm pairs}}$
(dashed line) and (b) the sum of $\rho^{{\rm B_{1g}}}_{{\rm pairs}}$
and $\rho^{{\rm B_{2g}}}_{{\rm pairs}}$ as a function of doping with
$T=0.002J$ are plotted in Fig. \ref{fig5} in comparison with the
corresponding experimental data \cite{blanc} (inset). The present
results show that in the momentum space, $\rho _{{\rm pairs}}$ is
strongly anisotropic as a function of doping. Although $\rho^{{\rm
B_{1g}}}_{{\rm pairs}}$ is much larger than $\rho^{{\rm B_{2g}}}
_{{\rm pairs}}$ in the corresponding doping, they almost linearly
increase with increasing doping in the underdoped regime. In
particular, at low doping, $\rho^{{\rm B_{1g}}}_{{\rm pairs}}$ and
$\rho^{{\rm B_{2g}}}_{{\rm pairs}}$ become very small and vanishe
below $p\sim 0.05$. Moreover, in a similar way to the superfluid
density, the sum of $\rho^{{\rm B_{1g}}}_{{\rm pairs}}$ and
$\rho^{{\rm B_{2g}}}_{{\rm pairs}}$, which describes the overall
DOCP, increases linearly with increasing doping in the underdoped
regime, in qualitative agreement with the experimental data
\cite{blanc}.

A nature question is why the doping and temperature dependence of
the $B_{1g}$ spectrum in the overdoped regime and $B_{2g}$ spectrum
throughout the SC dome in cuprate superconductors can be described
qualitatively within the KE driven SC mechanism. The reason is that
the KE driven SC state is the conventional d-wave BCS like
\cite{feng03,guo07}. In particular, this KE driven d-wave SC state
is controlled by both SC gap and quasiparticle coherence, then the
maximal $T_{c}$ (then gap parameter) occurs around the optimal
doping, and decreases in both underdoped and overdoped regimes. On
the other hand, based on the phenomenological BCS formalism with a
simple d-wave gap $\Delta_{\bf k}=\Delta [{\rm cos}k_{x} -{\rm cos}
k_{y}]$, it has been shown \cite{d1} that the positions of the
maxima in the $B_{1g}$ and $B_{2g}$ scattering both scale with the
gap parameter $\Delta$. However, in the present paper, the ERR
function of cuprate superconductors (\ref{ramanscattering}) is
obtained in terms of the imaginary part of RDDCF
(\ref{ramancorrelationfunction}) within the KE driven SC mechanism,
although its expression is similar to that obtained within the
phenomenological d-wave BCS formalism \cite{d1}. This is why the
doping and temperature dependence of the $B_{1g}$ spectrum in the
overdoped regime and $B_{2g}$ spectrum throughout the SC dome in
cuprate superconductors can be described qualitatively within the KE
driven SC mechanism.

Although many different theories have been put forward in order to
explain the nature of ERR in the $B_{1g}$ orientation in the
underdoped regime, however, its full understanding is still
controversial. One point of view is that the experimental result
from ERR \cite{e4,e5,e6,e7,e8,e9} does not necessary imply two
distinct gaps, but still it does imply that the gap behaves
differently in the nodal and antinodal momentum regimes. In
particular, it has been argued that the ERR data in both $B_{1g}$
and $B_{2g}$ geometries in the underdoped regime can be equally well
explained within a one-gap scenario if final-state interactions are
taken into account \cite{chubukov}. In other theories, the normal
state pseudogap is associated with the nature of the $B_{1g}$ peak
energy in the underdoped regime \cite{hufner08}. However, in the
present case, the continuation of the treatments from the SC to
normal states is not straightforward. This can be seen directly from
Eq. (\ref{ramancorrelationfunction1}), the intensity of ERR vanishes
proportional to $\bar{\Delta}^{2}_{Z}({\bf k})$ as $T\rightarrow
T_{c}$. In this case, ERR in the underdoped regime has been studied
within the fluctuation-exchange approximation for the Hubbard model
by phenomenologically introducing a pseudogap \cite{dahm99}, where
this pseudogap leads to that a corresponding peak evolves
continuously in the $B_{1g}$ spectrum as $T$ decreases in the normal
state and below $T_{c}$. However, this theory can not be applied to
discuss the doping dependent behaviors of ERR in $B_{1g}$ and
$B_{2g}$ symmetries.

Within the KE driven SC mechanism, our present result of the doping
and temperature dependence of ERR in the $B_{1g}$ symmetry is in
disagreement with the experimental data
\cite{e1,e2,e3,e4,e5,e6,e7,e8,e9,e10,blanc} in the underdoped
regime. In Ref. \onlinecite{e8}, it has been argued that for a
correct description of the $B_{1g}$ spectrum in the underdoped
regime, two essential ingredients should be taken into account
within the d-wave BCS formalism: (a) the quasiparticle spectral
weight $Z_{F}({\bf k})$ as well as the vertex correction; and (b) a
general gap also should be taken into account by including the
higher order of the harmonic component in the simple d-wave gap
function $\Delta_{\bf k}=\Delta [{\rm cos} k_{x}-{\rm cos} k_{y}]$.
In our present studies, the first condition is partially satisfied,
since the doping dependence of the quasiparticle spectral weight
$Z_{F}$ has been included in our above discussions. However, the
vertex correction for the interaction of the quasiparticles due to
the presence of the spin fluctuation is not included, which also is
doping and momentum dependent. For the second condition, a more
general gap $\Delta_{\bf k}=\Delta[{\rm cos}k_{x}-{\rm cos}k_{y}+
B({\rm cos}2k_{x}-{\rm cos}2k_{y})]$ has been observed
\cite{mesot99}. Although this gap function still is basically
consistent with the d-wave symmetry, it is obviously that there is a
significant deviation from the simple d-wave form. On the other
hand, the final-state interactions are also very important for
interpreting the unusual $B_{1g}$ spectrum in the underdoped regime
as point out in Ref. \onlinecite{chubukov}, and should be taken into
account. Thus an important issue is how these essential ingredients
that have been dropped in the present studies are taken into account
within the KE driven SC mechanism for a correct description of the
$B_{1g}$ spectrum in the underdoped regime. These and the related
issues are under investigation now.

In summary, we have studied the doping and temperature dependence of
ERR in cuprate superconductors based on the KE driven SC mechanism.
Our results show that the temperature dependent depletion at
low-energy shifts is faster in the $B_{1g}$ symmetry than in the
$B_{2g}$ symmetry. When increasing doping, the intensities of the
peaks in both $B_{1g}$ and $B_{2g}$ orientations are strongly
increased. In particular, in analogy to the domelike shape of the
doping dependence of  $T_{c}$, the maximal peak energy in the
$B_{2g}$ channel occurs around the optimal doping, and then
decreases in both underdoped and overdoped regimes. Moreover, in a
similar way to the superfluid density, the overall DOCP increases
with increasing doping in the underdoped regime.

\acknowledgments

This work was supported by the National Natural Science Foundation
of China under Grant Nos. 10774015 and 11074023, and the funds from
the Ministry of Science and Technology of China under Grant Nos.
2006CB601002 and 2006CB921300.

\end{document}